\documentclass[a4paper,11pt]{article}
\pdfoutput=1 

\usepackage[english]{babel}		
\usepackage{cite}				
\usepackage{graphicx}
\usepackage[utf8]{inputenc}		
\usepackage{jinstpub} 
\usepackage{lineno}				
\usepackage{mathtools}			
\usepackage{siunitx}
\usepackage{subcaption}

\usepackage[table]{xcolor}

\usepackage{arydshln}
\ADLinactivate

\DeclarePairedDelimiter{\ceil}{\lceil}{\rceil}
\DeclarePairedDelimiter{\floor}{\lfloor}{\rfloor}

\graphicspath{{Image_Resources/}}


\title{Implementation of high-performance, sub-microsecond deep neural networks on FPGAs for trigger applications}

\author{N. Nottbeck,}
\author{Dr. C. Schmitt,}
\author{Prof. Dr. V. Büscher}

\affiliation{Johannes Gutenberg-Universität Mainz}

\emailAdd{schmittc@uni-mainz.de}

\abstract{Deep neural networks are already widely used for physics analysis, but there are only few applications within low-level hardware triggers, and typically only with small networks. Modern high-end FPGAs offer Tera-scale arithmetic performance, and thereby provide a significant amount of operations per data set even for \si{MHz}-range data rates. We present a bottom-up approach of implementing typical neural network layers, in which we took into account both the special constraints that come from high-performance trigger systems, such as the ATLAS hardware trigger at the LHC, as well as an efficient implementation. By specifically designing each layer type to match our requirements, we could develop a framework that reaches 90 to 100\% processing efficiency for large layers, requires only few extra resources for data flow and controlling, and offers latencies in the range of only tens to hundreds of nanoseconds for entire (deep) networks. Additionally, a toolkit was built around these optimized layer implementations, which facilitates the creation of the FPGA implementation of a trained NN model.}


\begin{document}

\maketitle
\flushbottom

\section{Introduction}

Deep neural networks (DNNs) are already used in a wide range of inference tasks, such as speech and image recognition, and are continuously advancing into physics, until now mostly for offline data analysis. There are only very few examples where DNNs are used in or targeted for the context of high-performance detector triggers. This might be due to the very special inference rate and latency constraints that apply in such environments, and the difficulty of developing complex algorithms for field-programmable gate arrays (FPGAs). FPGAs, however, often are the only type of processing hardware that can be used in such contexts, apart from even more demanding ASICs. One example of a specific network in the trigger context can be found in the Belle II trigger, in which a relatively small neural network is used for z-vertex triggering~\cite{Neuhaus:2017trg}. An example for a more general attempt at enabling neural network usage within triggers is the "High Level Synthesis for Machine Learning" (hls4ml) companion compiler, which uses high-level synthesis tools to generate the FPGA firmware design for a given network \cite{Duarte:2018ite}. A more detailed overview on trigger requirements and existing work regarding neural network inference on FPGAs is also given in \cite{Duarte:2018ite}.
In this paper we take the ATLAS detector~\cite{PERF-2007-01} and its upgrades of the first level trigger system~\cite{PHASE1,PHASE2} as a reference for our studies. In this FPGA-based trigger level, the incoming data rate of \SI{40}{MHz} needs to be reduced down to less than \SI{100}{kHz} within a maximal latency of \SI{2.2}{\micro\second}. Only this large reduction in rate allows for the further processing by a software trigger that reduces the event rate further to about \SI{1}{kHz}, the maximum rate that can be written to permanent storage. Most of the \SI{2.2}{\micro\second} latency is used up by data preparation and transfer, such that only a few tens to few hundreds of nanoseconds remain for actual neural network applications.

In contrast to the hls4ml framework, we chose to pursue a hardware-centric, bottom-up approach for implementing general neural networks on FPGAs, which grants maximum control over the FPGA design, and therefore allows very fine tuning for the specific use case. Thereby, we intend to further lift network size limits, while simultaneously providing scientists working on trigger algorithms with an easy-to-use tool for efficiently incorporating neural networks of sizes that were never used before into their systems.

We begin with a very brief overview over the relevant aspects of neural networks as well as FPGAs. Following that, we discuss the required types of operations, how they can be implemented in hardware, and introduce the concept of user-configurable fixed-point precision. The main section of this paper focusses on the design of the individual layers, which currently include \emph{fully-connected layers}, as well as two-dimensional (multi-channeled) \emph{convolutions} and \emph{maximum pooling}, and concludes this with the implementation of activation functions. We then quickly discuss what needs to be considered for putting multiple layers together in a functioning network, before the presentation of implementation results on individual layers and entire networks. In the end, we summarize the results of our developments and give an outlook on possible future improvements.

\section{Basics}

\subsection{Neural networks}

In the following, we are focussing on deep neural networks which consist of fully-connected, 2D convolutional and 2D maximum pooling layers, and any meaningful combination of these, i.e. an arbitrary combination of the 2D layers, which might be followed by flattening and then an arbitrary sequence of fully-connected layers, or alternatively a network consisting of fully-connected layers only. As neural network framework, Keras was used, with a TensorFlow backend \cite{Keras}.

In a fully-connected layer, every neuron receives every input, with the number of inputs $N_\mathrm{I}$ being predetermined by the network and the number of neurons $N_\mathrm{N}$ being a parameter of the layer. The inputs $\vec{i} \in \mathbb{R}^{N_\mathrm{I}}$ are multiplied by a weight matrix $W \in \mathbb{R}^{N_\mathrm{N} \times N_\mathrm{I}}$, typically have an offset $\vec{b} \in \mathbb{R}^{N_\mathrm{N}}$ applied and then go through a usually component-wise real activation function $\vec{A}: \mathbb{R}^{N_\mathrm{N}} \rightarrow \mathbb{R}^{N_\mathrm{N}}$ to produce the result $\vec{o} \in \mathbb{R}^{N_N}$, i.e. the fully-connected layer implements equation \ref{eq:Dense}.
\begin{equation}\label{eq:Dense}
	\vec{o} = \vec{A}\left( W\cdot\vec{i} + \vec{b} \right)	
\end{equation}

A 2D convolutional layer receives an input image/feature map of shape $H_\mathrm{I} \times W_\mathrm{I} \times D_\mathrm{I}$. The number of kernels $N_\mathrm{K}$ and the kernel area $H_\mathrm{K} \times W_\mathrm{K}$ are parameters, while the kernel depth $D_\mathrm{K}$ equals $D_\mathrm{I}$ to incorporate all input channels into the kernel application. For a general layer, there are two additional architectural parameters called padding and stride. The stride determines the step lengths between kernel applications, and padding determines how kernel applications at the edges are treated. Currently, we support the 'default' stride of 1 in both directions and what is usually called padding 'valid', i.e. no incomplete kernel applications at the edges. Accordingly, output height and width are computed as $H_\mathrm{O} = H_\mathrm{I} - (H_\mathrm{K} - 1)$ and 
$W_\mathrm{O} = W_\mathrm{I} - (W_\mathrm{K} - 1)$, respectively, with the output depth $D_\mathrm{O} = N_\mathrm{K}$. The application of a kernel is similar to the fully-connected neuron operation, where $\vec{i}$ contains the inputs that are covered at the given position and $W$ depends on the kernel that is used.

As the 2D convolutional layer, a 2D pooling layer also receives a potentially multi-channeled input image. Instead of a kernel area, a pooling area $H_\mathrm{P} \times W_\mathrm{P}$ is defined, which is applied channel by channel and selects the maximum value within the given input range. As for the convolutional layer, currently only the default stride is supported, which in this case is $(H_\mathrm{P}, W_\mathrm{P})$ (i.e. non-overlapping pooling areas without free spaces in between). Padding options 'valid', 'same' and 'unchanged' are supported (see section \ref{ss:layers_pooling}). In consequence, the output shape is at least $\floor[\big]{ \frac{H_\mathrm{I}}{H_\mathrm{P}} } \times \floor[\big]{ \frac{W_\mathrm{I}}{W_\mathrm{P}} } \times D_\mathrm{I}$, with possibly one more element in height and width depending on the padding option and shapes.

Generally, it is also possible to reshape the data between layers. Until now, we support flattening from 2D multi-channeled data to 1D, where inputs are rearranged in a C-like manner, i.e. an input at $(x,y,channel)$  will be put into position $x\cdot (W_\mathrm{I} \cdot D_\mathrm{I}) + y\cdot D_\mathrm{I} + channel$ in 1D.

\subsection{FPGAs}
\label{Basics_FPGAs}

Modern FPGAs comprise a large number (up to the order of millions) of programmable look-up tables (LUTs) and typically one or two flip-flop registers (FFs) per LUT, with the routing between these also being progammable. In the following, any specifics apply to the Xilinx UltraScale+ (US+) FPGA architecture \cite{Xilinx_USplus, Xilinx_USplus_SwChar, Xilinx_CLB, Xilinx_DSP}. However, most features are directly or similarly applicable to other recent device families from Xilinx and competitors such as Intel/Altera.

The LUTs typically implement binary combinational functions from few bits to 1- or 2-bit results, e.g. $f:\{0, 1\}^6 \rightarrow \{0, 1\}^2$ in the US+ architecture.\footnote{True di-output LUTs are only supported for up to 5 inputs, otherwise constraints apply, see \cite{Xilinx_CLB}.} A flip-flop can store a single bit.

The combination of a large ensemble of programmable logic functions, registers for storage and synchronous processing and programmable routing makes it possible to implement even complex and large digital circuits within FPGAs.

In addition to these basic building blocks, there are also specialized embedded components such as block memories (BRAMs) and digital signal processors (DSPs). The BRAMs provide a high-density, high-capacity memory. One BRAM block consists of two \SI{18}{kib} parts, where each is addressed by 9 to 14 address bits, which results in port widths and depths ranging from 36 and 512 to 1 and 16384. The DSPs are 'simple' ALUs (arithmetic logic units), which offer a wide range of operations. In the US+ architecture, the operation mode which is most important for NN applications is a $w \cdot i + b$ type operation, where $w$, $i$ and $b$ are binary numbers of up to 18, 27, and 48 bits, respectively. The number $b$ can be an external input or come from an internal result register, allowing to perform a multiply-accumulate operation in either a pipelined (if $b$ is external) or localized (if $b$ is internally accumulated) design.

Our target device was the Xilinx US+ XCVU9P, which features approximately \SI{1.2}{M} LUTs and \SI{2.4}{M} FFs, as well as 6840 DSPs and 2160 2x18\,kib BRAM units, and supports maximum operation frequencies in the range from  \SI{640}{MHz} to \SI{900}{MHz}, depending on device speed grade and hardware component. The FPGA firmware was designed in VHDL using Vivado 2018.2 as IDE with default settings for synthesis and implementation strategy, except for the synthesis mode \emph{out of context}, to make it possible to implement designs without any I/O connections. Apart from automatic inference of hardware components such as LUTs, FFs and DSPs from the VHDL code, they were also explicitly instantiated by \emph{design primitives} for the DSPs and adders implemented in the general logic, which gave maximum control over the implementation.

Key metrics for the evaluation of an FPGA implementation are the amount of resources required and the design timing, i.e. if the target frequency is met or if the design cannot run at the targeted frequency due to time needed for signal propagation. Designs which miss timing cannot be used without optimization, but even in these cases, it is useful to have knowledge about the severity of the timing violation.

\subsection{Arithmetics implementation}

\paragraph{Precision requirements}
Previous work~\cite{Duarte:2018ite} has demonstrated that for NN inference, reasonably low precision is sufficient, and neither floating-point arithmetics nor a large number of bits are necessary for optimal performance in many cases.\footnote{Which we also verified for our own test implementations.} Using fixed-point arithmetics provides the benefit of a significantly reduced implementation complexity, which saves resources and makes the adjustment of the arithmetic precision to specific needs regarding value range and granularity significantly easier.

In the following, we define the precision specification '$i.f$', where $i$ and $f$ are the integer and fractional bits in a fixed-point representation, with the entire value being in the so called \emph{two's~complement} representation. Accordingly, the resulting value range is $- 2^{i - 1}$ to $2^{i - 1} - G$ with the granularity $G = 2^{-f}$.

\paragraph{Implementation in hardware}

For DNNs as described above, only a few basic operations are needed: Multiplication and addition are combined into the multiply-accumulate (MAC) operation, and for the pooling layers, the 'select maximum' operation (MAX) on two values is required.

Since fixed-point arithmetics are sufficient, it is possible to use only a single DSP to implement the MAC operation that is required for computing weighted input sums.\footnote{This of course limits the number of bits for the inputs, but with up to 27 by 18 bit-wide multiplications in the UltraScale+ architecture, this limit is significantly above what is usually critical for neural network purposes.} Thereby, it is possible to provide one input and weight to a DSP per cycle, and then the DSP can add the product either to the internally accumulated value or to an externally provided partial result. For reaching the maximum DSP frequency, it is necessary to use internal pipeline registers of the DSPs. This requires having internal registers enabled at three levels: input, product and after the accumulation. Due to this, a 'first DSP' latency of three cycles arises. For any additional DSP in a pipeline, only one extra cycle is necessary, as even an externally provided extra term (e.g. the current partial result) can be inserted latency-free before the accumulation step.\footnote{An extra external input register is also possible and might be required for future frequency improvements, but it is possible to incorporate this with only one extra cycle for the entire DSP pipeline, instead of one cycle per DSP.}

In some situations, it is necessary to compute sums of values that come out of multiple DSP pipelines. Adding $N$ values requires $N-1$ adders. By adding multiple values in a binary tree structure, where at each level, pairs of values are added and the result is sent to the next level, it is possible to keep the logic depth at the optimal value of $\ceil[\big]{ \log_2 N }$, which is desirable from a latency point of view. To guarantee optimal performance, these logic-based adders were instantiated on a design primitive level, forcing Vivado to use the dedicated carry logic and to place the corresponding primitives as close as possible.

For the pooling layers, it is necessary to find the maximum in a set of values. Similarly to the sum of multiple values, this can be done in a binary tree structure, to provide the lowest possible logic depth and resource utilization.\footnote{It might be possible to use an algorithm with formally lower logic depth, but this is not expected to scale to the hardware regarding utilization and timing.} Instead of being added, each pair of values is processed by the MAX operation, to detect the larger and send it to the next level, until only one value remains.

For the maximum of and sum of set structures, we added configuration parameters to include register stages at the input, inter-operation and output levels, allowing to trade maximum frequency against register utilization and latency in terms of cycles.

\section{Layer structure}

\subsection{General decisions}

Due to the very strict real-time context of high-performance detector triggers, it is very important to optimize the design for minimum latency.
At the same time, it is necessary to aim for an optimized inference performance, which strongly relates to an efficient resource usage. It is not expected that a solution could be found which optimizes all three metrics at once, especially if no further constraints exist for the network architectures. The selection of potentially suitable design approaches is tightly bound to the data rate $f_\mathrm{D} = T_\mathrm{D}^{-1}$ (with the data tick period $T_\mathrm{D}$), the achievable processing frequency $f_\mathrm{P}$ and the latency budget. Motivated by the ATLAS first level trigger, we chose to optimize the design for a data rate of \SI{40}{MHz} and for a latency as short as few tens of nanoseconds for entire networks, but put performance and efficiency before a perfectly minimized latency.

Generally, if data is coming in at a rate $f_D$ and the frequency of the processing units (PUs), e.g. the DSPs, is $f_\mathrm{P}$, there are $C = \floor[\big]{ \frac{f_\mathrm{P}}{f_\mathrm{D}} }$ cycles per PU and data set for real-time processing. Moreover, three general structures could be used that potentially reach perfect efficiency: Ideally, one would have all of the hardware resources available to process an entire data set within one data tick, and then proceed with the next set, which also results in the shortest latency. If that is not possible, one can either build a single pipeline of processing steps, where partial results of one pipeline stage propagate to the next stage during each data tick, or partition the hardware into $N_\mathrm{P}$ parts, and place the corresponding number of instances of the network processors, where each has $N_\mathrm{P}$ data ticks for processing, and inputs are multiplexed to a different instance at each tick. We have decided for a single-instance pipelined design, and excluded the others:

The 'single tick processing' design can be ruled out because with e.g. $T_\mathrm{D} = \SI{25}{ns}$ for the ATLAS detector, it will not be possible to have large designs which process all of the data within one data tick, simply due to the time required for computation and signal propagation in device-spanning, deep logic structures.

The decision for either a single pipeline design or a multi-instance design is less trivial, as both have advantages: In the single pipeline design, every pipeline stage could potentially be adapted to a specific task, which would reduce overhead structures. A multi-instance design would require more general processing units instead (as these would need to perform different tasks/compute different layers while time progresses), and could be implemented for example as systolic arrays. At the same time, data propagation through the single pipeline design would be bound to the pipeline structure, while the processing units in the multi-instance design could \emph{potentially} allow better adjustment of the latency depending on the exact design, although at the price of an increased complexity.

We took three further aspects into account for our decision: In order to avoid an increase in latency, we decided not to batch the input data, i.e. the batch size during inference is 1. Additionally, with the given maximum hardware processing frequency and targeted data frequency range, $C$ is in the order of $10^1$, and in a neural network, a layer can only begin computing when at least some of the required inputs are available. 

All of this convinced us to choose the single pipeline design: It makes it easier to keep the efficiency high if no batching is used and $C$ is small, it allows having relatively small (although possibly not minimized) latencies and it avoids too much overhead structure for reconfigurability and data flow management.\footnote{Even with this design, it is possible to further reduce the latency at cost of computational efficiency by constraining the layers to use less than $C$ cycles. The loss in computational efficiency would typically be tolerable, as mostly small networks would benefit from this, while larger networks are rather throughput-bound than latency-bound anyway.} For the pipelining approach, it also appears natural to have one pipeline stage per network layer, which is conceptually much less complex than making efficient use of a systolic array while simultaneously keeping the latency low. Layers were therefore designed to take the data within $C$ cycles after the first part of the data arrives, and also produce all results within a period of $C$ cycles, to maintain network synchronicity, with an arbitrary but fixed delay in between.

\begin{figure}
	\includegraphics[width = \textwidth]{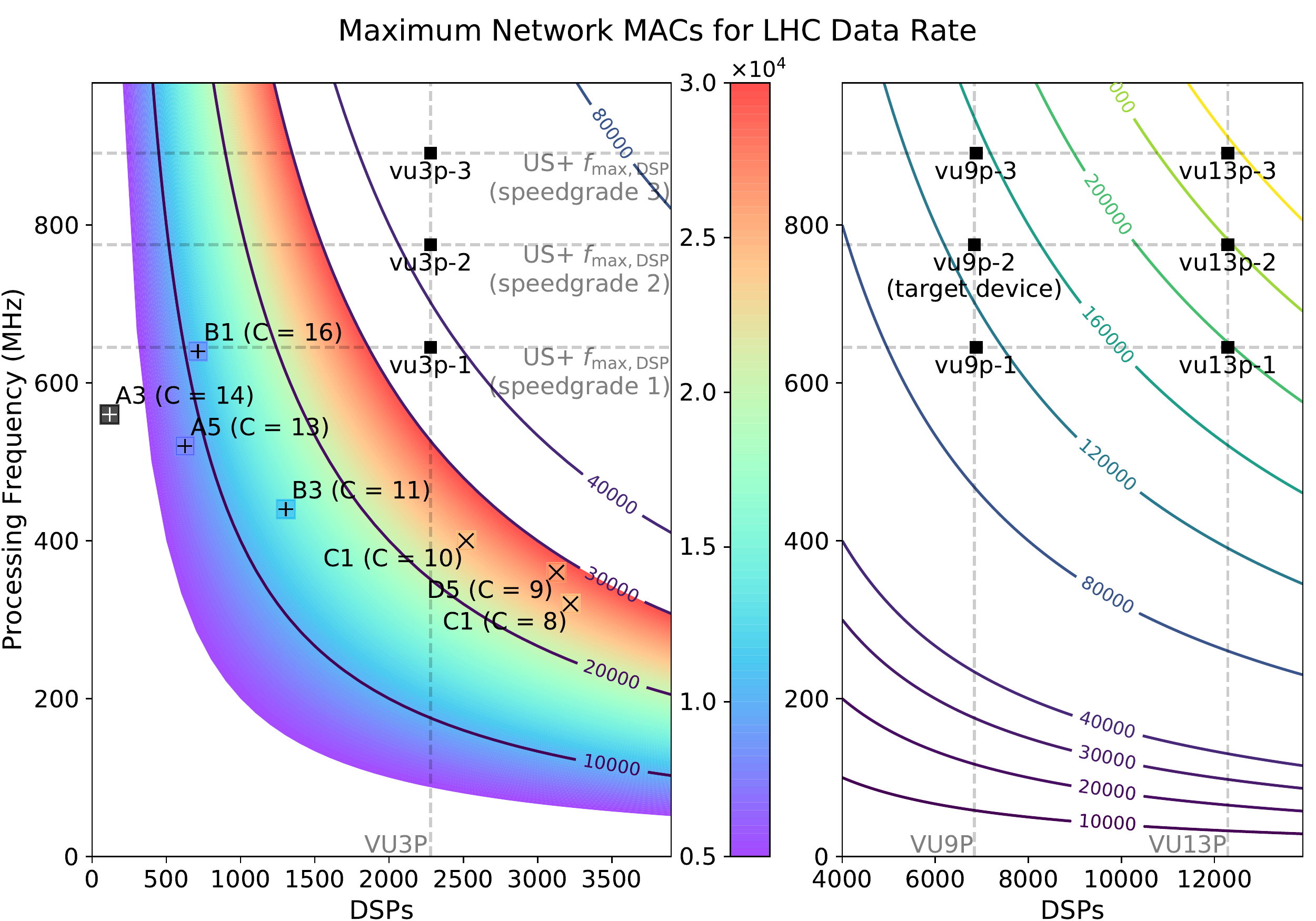}
	\caption{\label{f:TOPsVsDevice}
          Maximum network size (in terms of MACs per data set) depending on processing frequency and number of DSPs, together with specifications of devices from the Xilinx UltraScale+ family (dashed grey lines and black squares). The number of operations is obtained by dividing the product of frequency and DSPs by a data frequency of $f_\mathrm{D} = \SI{40}{MHz}$. Network implementations which met the timing are marked with a plus, other examples with a cross.
The color of the square in the marker background indicates where the network would be located if all processing cycles could be used, i.e. it is an indicator for the ratio of used cycles (no contrast corresponds to a 100\% efficiency). (See section \ref{ss:Results_Networks} for details on these networks).
	}
\end{figure}

Moreover, with $N_\mathrm{PU}$ processing units placeable in a device, there are $N_\mathrm{Op} = C \cdot N_\mathrm{PU}$ operations possible per data set and operation type, which is useful to understand the ideal-case inference performance of any device and thereby the maximum network size. Figure \ref{f:TOPsVsDevice} shows the number of DSP MACs possible depending on the device/DSP count and processing frequency, where the ATLAS data frequency of \SI{40}{MHz} was assumed. This shows that already with one of the current high-end FPGA families, a network size of up to 280 kMACs could be possible even for such high data frequencies, if the maximum device frequency can be achieved and the computational efficiency is close to 100\%.

\subsection{Fully-connected layers}

\begin{figure}
	
	\centering
	
	\begin{subfigure}[b]{.85\textwidth}
		\includegraphics[width = \textwidth]{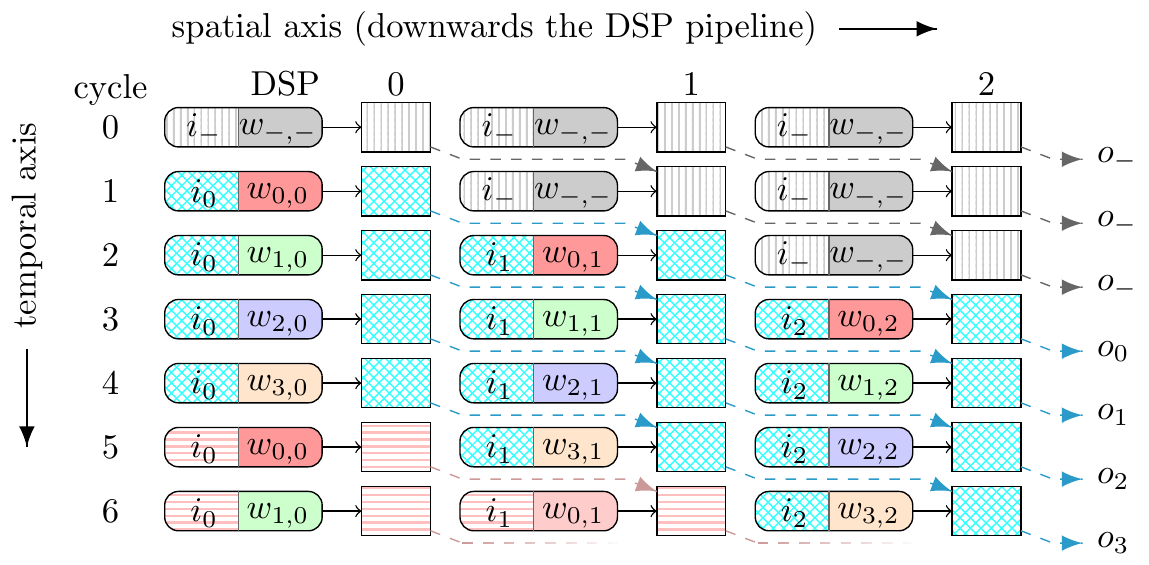}
		\caption{\label{fs:Dense_Dataflow}
                  Data flow schematic for the fully-connected layer. 
                  The color (pattern) coding is as follows: gray (vertical lines) refers to no data set available/idling, while cyan (crosshatch) and pink (horizontal lines) refer to data of a first and second data set, respectively. The inputs of neighboring pipeline stages are updated in subsequent cycles and only once per data cycle, while the weight sequence for each individual DSP is repeated during each data cycle.
		}
	\end{subfigure}
	\vspace{1\baselineskip}
	
	\begin{subfigure}[b]{.85\textwidth}
		\includegraphics[width = \textwidth]{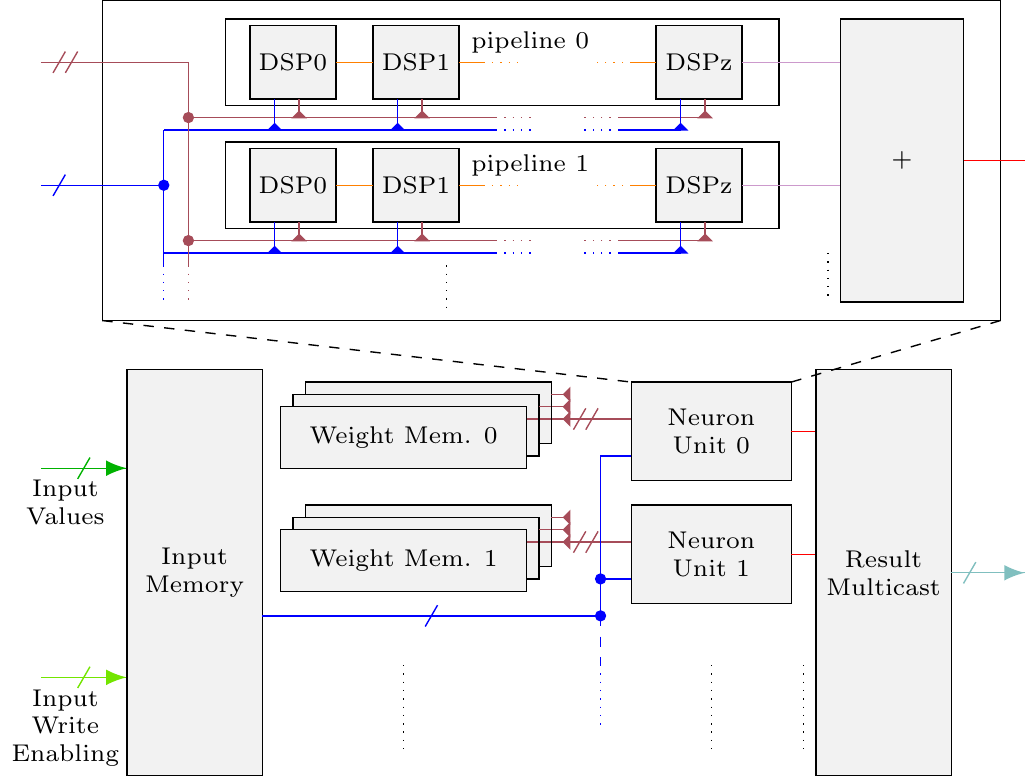}
		\caption{\label{fs:Dense_Structural}
			General structural schematic of a fully-connected layer. For connections of inputs and weights to DSPs see text. As in the following, the dimensionality of signals is indicated by the number of strokes on a given signal line. (No stroke corresponds to scalar data, one stroke to an 1D array, etc.)
		}
	\end{subfigure}
	
	\caption{Fully-connected layer design.}
	
\end{figure}

Generally, it is possible to use a single DSP to implement the weighted input sum of a neuron, as the DSPs support the MAC operation. However, the number of processing cycles is limited, therefore arbitrary amounts of inputs $N_\mathrm{I}$ could only be handled if multiple inputs are weighted in parallel in multiple DSPs, and even then (or if $N_\mathrm{I} < C$), such an approach can be computationally very inefficient.

Instead, we exploited that neurons in fully-connected layers require each of the inputs in order to be computed. By combining this with the inputs being made available within $C$ cycles, we implemented the weighted input sums using pipelines of DSPs.

A typical pipelined data flow is illustrated in figure \ref{fs:Dense_Dataflow}, where for illustrational purposes we assumed that per processing cycle, only one further input becomes available. In this scheme, there are as many DSPs as inputs to the given pipeline. The input $i_n$ ($n \in \mathbb{N}_0$), which is available at cycle $n$, is stored in a register which is connected to the input of $\mathrm{DSP}_n$ at the subsequent cycle edge, which keeps this input until it is overwritten after $C$ cycles. With $n + 1$ being the first cycle when $\mathrm{DSP}_n$ has the new input available in its register, weight $w_{m,n}$  is presented to $\mathrm{DSP}_n$ during cycle $n + 1 + m$, where $m \in \mathbb{N}_0$ denotes the neuron index. After each cycle, the partially accumulated weighted input sum is passed as third input to the next DSP in the pipeline, which adds the next weighted input for the corresponding neuron, until the final result comes out of the last DSP.

Multiple of these pipelines can be used when multiple inputs become available during each cycle. We define the number of pipelines as $P$. Each pipeline weights at least $\floor[\big]{ \frac{N_\mathrm{I}}{P} }$ inputs, and the first $N_\mathrm{I} \bmod P$ pipelines need to weight one extra input, while the remaining pipelines are extended by a shift register for cycle-matching the partial results.

Apart from parallelizing the pipelines when multiple inputs become available per processing cycle, it is also possible to use multiple of such 'neuron units' (NUs) if necessary, i.e. $N_\mathrm{NU} = \ceil[\big]{ \frac{N_\mathrm{N}}{C} } > 1$. In that case, each NU computes at least $\floor[\big]{ \frac{N_\mathrm{N}}{N_\mathrm{NU}} }$ neuron results, and the first $N_\mathrm{N} \bmod N_\mathrm{NU}$ NUs compute one extra neuron. If $\ceil[\big]{ \frac{N_\mathrm{N}}{N_\mathrm{NU}} } < C$, all neuron results are available even before $C - 1$ extra cycles after the first result, which allows a reduction of the layer-to-layer latency.

\paragraph{Code structure}
As for the following layers, the fully-connected layer consists of multiple design subunits, which are shown in figure \ref{fs:Dense_Structural}. The main part of the fully-connected layer are the neuron units. These contain the DSP pipelines, which use the $w\cdot i + b$ type operation of the DSPs, with $w$ for weights, $i$ for the input values, and $b$ as partial result of the previous DSP. In the current implementation, the $b$ signal can be implemented either via dedicated connections between the DSPs or through the general fabric wiring. The latter case provides more placement flexibility, but worse timing characteristics. For the fabric routing, timing could be further improved by also activating the DSP input register for the $b$ signal, which was not yet implemented.\footnote{Minor adaptations in the layer structure and controlling are necessary for that.} If multiple pipelines are used, a final adder is included to sum up all partial results to produce the actual weighted neuron input sum.

The input values for the DSPs are stored in a memory entity with external write control, therefore the external write sequence must match the expected input sequence. Input $i$ is weighted by DSP $\floor[\big]{ \frac{i}{P} }$ in pipeline $i\bmod P$. Similarly, there are multiple weight memory blocks, realized with BRAM.\footnote{The choice for BRAM-based memory units fell because we considered these a spare and available resource, which makes it possible to save LUTs and registers. If necessary, it would be possible to switch to memory based on the general logic resources.} Currently, there is one weight memory block per NU and pipeline stage, which spans all DSPs of that stage and NU. In the future, other architectures might be added, such as blocks per pipeline or per part of a pipeline, for frequency optimization for different layer architectures.

In the end the results are multicast, i.e. the single result signal of any given neuron unit is then connected to all output signals of the entire dense layer which are produced by that neuron unit during any cycle.  A controller entity controls the selection of the weight memory and asserts signals indicating when a result is valid.

\subsection{2D convolutional layers}

Convolutional layers typically feature only few parameters, but a significant amount of the total MAC operations within a network, due to the manifold application of the same kernel. Obviously, it is possible to apply kernels at multiple positions in parallel, as there are no direct dependencies between different output positions.\footnote{There are partially shared inputs, however, and some very intelligent ways to reduce the computational cost, e.g. by making use of the convolution theorem to implement the convolution as multiplication, but these often work well only for larger layers than possible here and sometimes require preparations that introduce significant additional latency, therefore we did not focus on these.} However, not exploiting the topology of convolutional layers in any way results in a significant and mostly avoidable logic resource consumption, because the same input values and weights need to be copied often and in many places in spite of their redundancy.

Therefore, we studied possibilities to save resources by spatially and temporally reusing the layer input values as well as the weights. In the most extreme case, one would simply compute the entire result in parallel, which would mean that the weights could be completely static and every input would also only need to be stored (although dynamically updated per data set) exactly once. However, in such an architecture, most of the processing cycles would remain unused, which would result in an extremely inefficient resource usage, even though the data reuse is maximized. In consequence, it is necessary to find a way to divide the computations that reuses the data efficiently while also taking into account how many cycles are there for processing and that is feasible from an engineering point of view.

As example, computing an entire output channel at once would also allow an extremely efficient weight reuse, as all processing units could get the same set of weights for the current channel, and the weights would need to change only cycle by cycle. The input reuse would even still be maximized, as every output channel requires the complete input data. However, such a scheme would work well only if the number of processing cycles matches the number of output channels well, and otherwise could become extremely inefficient.

As other extreme, one could make use of nearly all of the available processing cycles if the computations are divided on a per-output-element basis, as for the fully-connected layer. While feasible in principle, this would make it very complicated to find and implement an efficient way of sharing inputs and weights between processing elements.

\begin{figure}
	\centering
	\includegraphics[width=.6\textwidth]{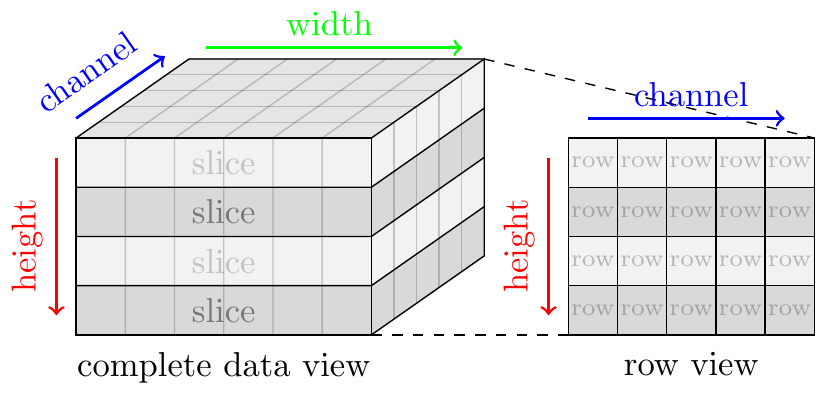}
	\caption{\label{f:slicing}
		Illustration of how 'slices' and 'rows' are related to the original data volume. A slice contains all elements at a given height, a row spans the entire width at a given height and channel position.
	}
\end{figure}

As compromise between the optimization of the data sharing and cycle usage, we decided to use one 'row' of data as smallest unit for distribution to different processing units, which we accordingly name 'row units' (RUs). We identify a row by its height and channel index, and it spans the entire width of the volume, see figure \ref{f:slicing}.

Generally, with $N_\mathrm{OE}$ output elements to be computed and $C$ cycles for processing, one requires at least $N_\mathrm{PU} = \ceil[\big] { \frac{N_\mathrm{OE}}{C} }$ processing units. With a row-wise division, there are $N_\mathrm{OE} = H_\mathrm{O} \cdot D_\mathrm{O}$ output elements to be computed. This means that $N_\mathrm{OE}$ is easily in the order of one hundred, while $C$ is expected to be 20 or significantly smaller in case of the ATLAS detector trigger. If we use these values as assumption for 'large' convolutional layers, the cycle efficiency (i.e. ratio of non-idling row unit cycles) is guaranteed to be at least 80\% even in the worst case, and up to 100\% in the best, which further improves with lower $C$ or larger layers. This is still not as good as what could be obtained for a per-element division, but it turns out that even for a row-wise division, one already needs to resort to a relatively complex design for combining the high cycle efficiency with an extensive data reuse, which is explained in the following.

The first level of data reuse is facilitated by the row unit itself: Neighboring output elements share a range of their inputs, and all of their weights. Accordingly, it is only necessary to provide one set of weights for the subunits within a row unit that compute the individual output elements, and only one range of the input data which is then shared internally.

A second level of data reuse can be introduced between row units: If row units are guaranteed to always process output rows belonging to the same channels in parallel, then it is also possible to let them share their weights throughout the entire computation.

Furthermore, it is also possible to use input sharing between the $N_\mathrm{RU} = \ceil[big] { \frac{H_\mathrm{O} \cdot D_\mathrm{O}}{C} }$ row units. For that, we want to introduce the concept of a 'slice', which contains all rows at a given height index, i.e. it is identified by the height index and spans the entire width and depth of the data volume, see figure \ref{f:slicing}. Row units that process rows from neighboring slices can share part of their inputs, as neighboring output slices require $H_\mathrm{K} - 1$ common input slices for their results.

Until here, there is 'for-free' data sharing within a row unit, and there are known conditions for inter-row-unit data sharing. Both belong to the spatial data reuse rather than the temporal data reuse. As final step, it is necessary to find a way to assign different output rows to row unit processing cycles, such that both the inter-row spatial conditions are met as well as possible and that temporal data reuse is also facilitated.

We developed the following scheme: Initially, all row units begin processing data for the same output channel, but of directly neighboring slices, i.e. row unit $i$ begins computing the result of the first channel of output slice $i$. Cycle by cycle, every row unit progresses to the next channel of the same slice. All row units begin computing outputs for the next free range of slices when all rows within the current slice range have been covered.

This scheme has the advantage that it maximizes both the input and the weight sharing between row units, and additionally, it is only necessary to load a new range of inputs when the row units move on to the next range of output slices, i.e. only once every $D_\mathrm{O}$ cycles. This means a drastic reduction of the amount of different inputs that need to be loadable, and therefore is very beneficial for the resource utilization, as this effects both registers for data storage and LUTs for input selection.

\begin{figure}
	
  \begin{subfigure}[b]{.35\textwidth}
    \centering
		\includegraphics[width=0.65\textwidth]{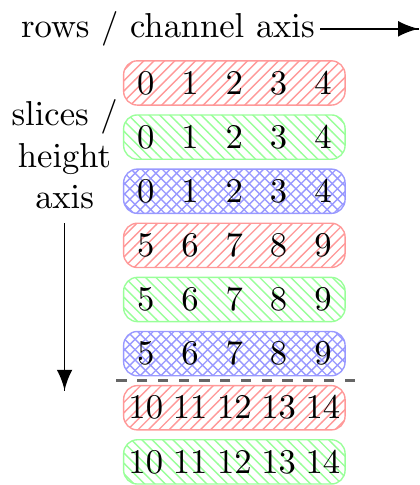}
		\caption{\label{fs:Conv_Allocs}
			Example of the row unit allocation scheme (\mbox{$C \in \lbrace 14,...,19 \rbrace$} and given output shape). Positions marked with a 0 are computed first, all others require the indicated number of delay cycles. The different colors (patterns) indicate which row unit covers a given row.
			The blue (crosshatched) row unit is 'short', the others 'long'.
		}
	\end{subfigure}
	\hfill
	\begin{subfigure}[b]{.62\textwidth}
		\includegraphics[width=\textwidth]{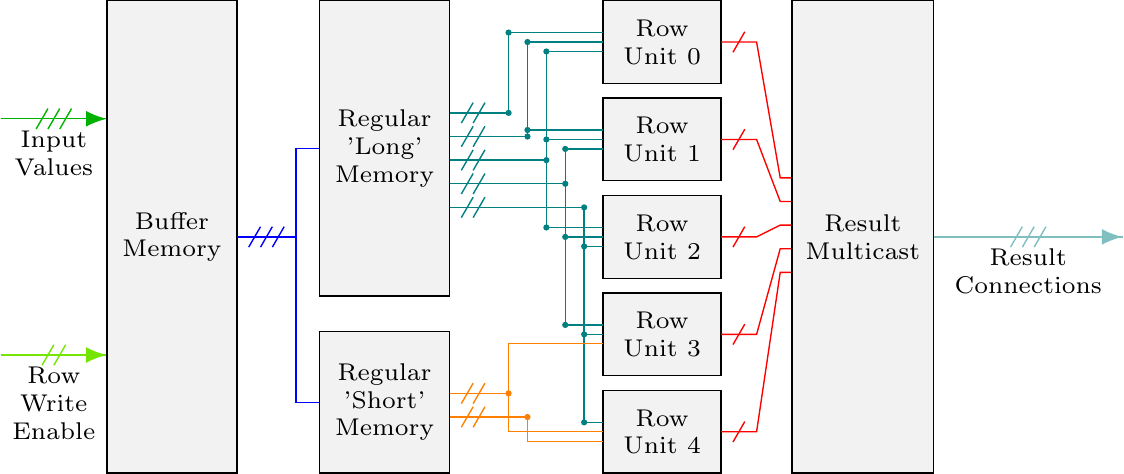}
		\caption{\label{fs:Conv_Structure}
			Convolutional layer schematic for the regular case (here with 5 RUs of kernel height 3, 3 of them 'long'). A buffer memory receives inputs and row-wise write enable signals. Internally, the row units receive their inputs from working memories, which provide selected ranges of the complete input. Finally, the RU results are multicast. Weight blocks for the row units and control infrastructure were left out for clarity reasons.
		}
	\end{subfigure}
	
	\caption{2D convolutional layer design (regular structure).}
	
\end{figure}

An example for this scheme is shown in figure~\ref{fs:Conv_Allocs}. Generally, any row unit can process at most $C$ rows, which can be part of $N_\mathrm{RU,Sl} = \floor[\big]{ \frac{C}{D_\mathrm{O}} }$ complete slices and $N_\mathrm{RU,Rem} = C \bmod D_\mathrm{O}$ remaining rows.

At this point, it is necessary to distinguish between a regular and an irregular case: In the regular case, there are at most as many slices as can be completely processed by the given number of row units, i.e. $N_\mathrm{Sl} \leq N_\mathrm{RU} \cdot N_\mathrm{RU,Sl}$. This means that it is possible to process all results for any slice by only a single RU. It might be necessary to have some RUs process one slice more than others, we attribute those which process an extra slice as 'long', the others as 'short'.

In the 'irregular' case, it is necessary to also use at least some of the $N_\mathrm{RU,Rem}$ remainder cycles of some row units, and the allocation scheme becomes much more complicated (see section \ref{ss:Conv_Irreg}).

\paragraph{Code structure}

\begin{figure}
	\centering
	\includegraphics{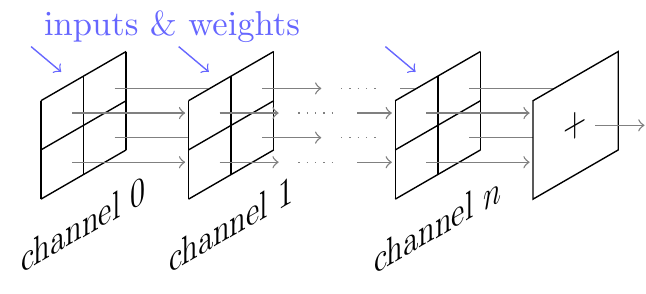}
	\caption{
		\label{f:Conv_Pipeline}
		2D convolutional layer pipeline structure. Inputs from the topmost channel are weighted first, the weighting of inputs then propagates towards the last channel cycle by cycle. Weighting and accumulation happens per input area element individually, and the total result is finally created by adding the partial results for all input area elements. Such a pipeline structure is used $W_\mathrm{O}$ times within a row unit, i.e. once for each row output position. Here illustrated for an $n$ channel deep, $2 \times 2$ kernel.}
\end{figure}

The main part of the 2D convolutional layer are the row units. These are provided with all of the weights and inputs they need. For a row with $W_\mathrm{O}$ output positions, there are just as many subunits for computing the respective output values. Every subunit is structured as in figure~\ref{f:Conv_Pipeline}: The inputs that are relevant for the given output position are weighted and accumulated per input channel from top to bottom channel, and partial results are pipelined individually for each kernel position. At the end, the total result for the given output position is produced by adding the partial results from all kernel positions. The top-to-bottom input scheme was used to adapt to the output scheme, which primarily also works from top to bottom channel, in order to typically allow faster data propagation between layers.

The rest of the layer was designed in an attempt to reduce the resource utilization (rather than frequency). A simplified structural example for the regular case is shown in figure~\ref{fs:Conv_Structure}. There is one input buffer memory, which has external write controlling and stores each of the (2D multi-channeled) inputs exactly once. The buffer memory is followed by working memories, which have control inputs for selecting when data is written to them and from which range of inputs data is taken. There always is a 'long', and if 'short' RUs are present, also a 'short' working memory. The 'long' working memory contains all input slices required for the 'long' RUs. The 'short' working memory contains all \emph{extra} input slices required for the 'short' RUs, i.e. those which are not required in any 'long' RU. We chose to differentiate the working memories into these two subunits because the 'short' memory needs to load one set of inputs less from the buffer memory, as the 'short' RUs finish their computation one cycle earlier, and therefore this structure can reduce the resource utilization. In figure \ref{fs:Conv_Structure}, each working memory output signal corresponds to an input slice, i.e. this spans the entire width and all channels, but has a fixed height index (which varies over time, while different ranges of inputs are loaded into the working memories).

To save further resources, we grouped RUs which always process the same output channel during any given cycle together, and gave all of them only a single weight memory. In the regular case, this means that there is only one set of weight memories for all 'long' and one set for all 'short' RUs. Details on the more complicated irregular case are described in appendix~\ref{ss:Conv_Irreg}.

As for the fully-connected layer, the row unit results are finally connected to all output positions where they produce results during any cycle (see allocation scheme), and the output row write enable signal is managed by a controller, which also controls the working memory input selection and write enabling and the weight memories.

\subsection{2D max-pooling layers}
\label{ss:layers_pooling}

\begin{figure}
	
  \begin{subfigure}[b]{.36\textwidth}
    \centering
		\includegraphics[width = 0.65\textwidth]{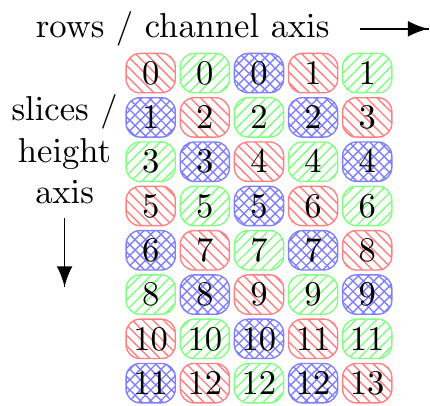}
		\caption{\label{fs:Pool_Allocs}
                  Example of the row unit allocation scheme (\mbox{$C \in \lbrace 14,...,19 \rbrace$} and given output shape). Positions marked with a 0 are computed first, all others require the indicated number of delay cycles. The different colors (patterns) indicate which row unit covers a given row.
		}
	\end{subfigure}
	\hfill
	\begin{subfigure}[b]{.61\textwidth}
		\includegraphics[width = \textwidth]{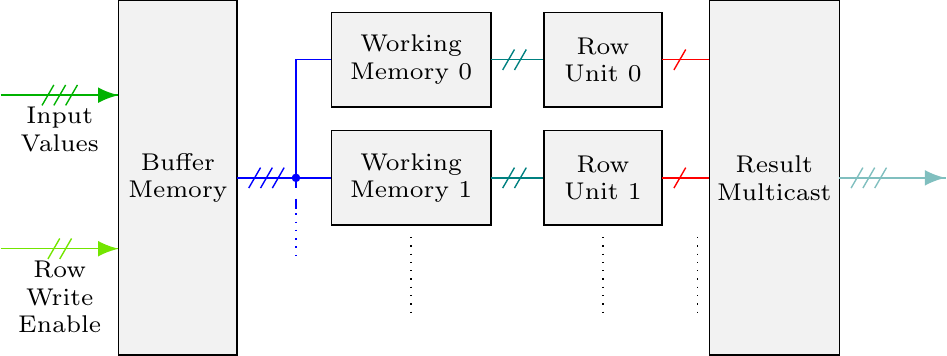}
		\caption{\label{fs:Pool_Structure}
			Pooling layer schematic. A buffer memory receives inputs and row-wise write enable signals. Internally, the pooling row units receive their inputs from working memories, which provide selected input rows. Finally, results are multicast. Control infrastructure was left out for clarity reasons.
		}
	\end{subfigure}
	
	\caption{2D maximum pooling layer design.}
	
\end{figure}

For default-stride pooling layers, where there is no input sharing, it is neither necessary nor possible to resort to as elaborate schemes as for the convolutional layers. Since all inputs are required only once, there is no way of saving resources or input accesses, and there is no need to use complicated row allocation patterns. For simplicity reasons, the concept of output rows and row units was still maintained, but the row allocation was simply done from top to bottom channel, top to bottom slice in an interleaved manner (see figure \ref{fs:Pool_Allocs}).

Contrary to the convolutional layers, it was already possible to implement different padding options for the pooling layer: These include padding 'valid' and 'same' (as in Keras), and a new mode 'unchanged'.\footnote{'Valid' padding dismisses 'incomplete' pooling inputs at the high-index edges, 'same' symmetrically extends the input space to only have complete pooling inputs, 'unchanged' extends only on the high-index edges.} Internally, padding is realized by the extension or truncation of the input signals. The latter will not increase the utilization significantly, as constant propagation will be detected and the corresponding logic parts removed/simplified by Vivado.

\paragraph{Code structure}
As for the convolution layer, the basic units of this layer are the row units. A row unit gets all inputs required to compute a single output row. Since pooling usually does not extend over multiple channels, this means that a row unit only requires a range of input rows of a single channel for each cycle.

A structural schematic of the design is shown in figure \ref{fs:Pool_Structure}. Similar to the convolutional layers, a buffer memory is used, from which the row unit working memories loads a new input slice during each cycle. This process is driven by a controller entity, which also controls the output write enabling. Row unit results are again multicast to any position in which they are needed at some cycle.

\subsection{Activation functions}

Apart from the obvious linear/identity activation, we currently support the \emph{rectified linear unit} (relu) activation, which is quite commonly used in (convolutional) deep neural networks, and has the advantage that it can be implemented with very little cost on FPGAs.

By explicitly instantiating on a design primitive level, we were able to guarantee a relu implementation that requires either $\floor[\big]{ \frac{B}{2} }$ LUTs or $B - 1$ FFs per relu unit working on $B$ bit values. For the LUT-based implementation, we used that Xilinx UltraScale+ LUTs support dual-output use for up to five common inputs. The relu is then obtained by providing two input value bits and the input value sign bit to a LUT, which either replicates the two input bits on its outputs (positive sign) or sets the outputs to zero (negative sign). The FF-based implementation makes use of the FDRE primitive\footnote{This realizes a D-type flip-flop with a synchronous reset and a clock enable input, the latter was tied to a logical '1'.} for every non-sign input value bit. The flip-flops simply store the input value bits if the sign is positive and reset to '0' if the sign is negative, i.e. the sign bit acts as reset control bit. This is an elegant demonstration of how advanced storage primitives can be used to implement not only storage, but also simple computations, which can save LUT resources for more complex operations.

\begin{figure}
	\centering
	\includegraphics[width = .8\textwidth]{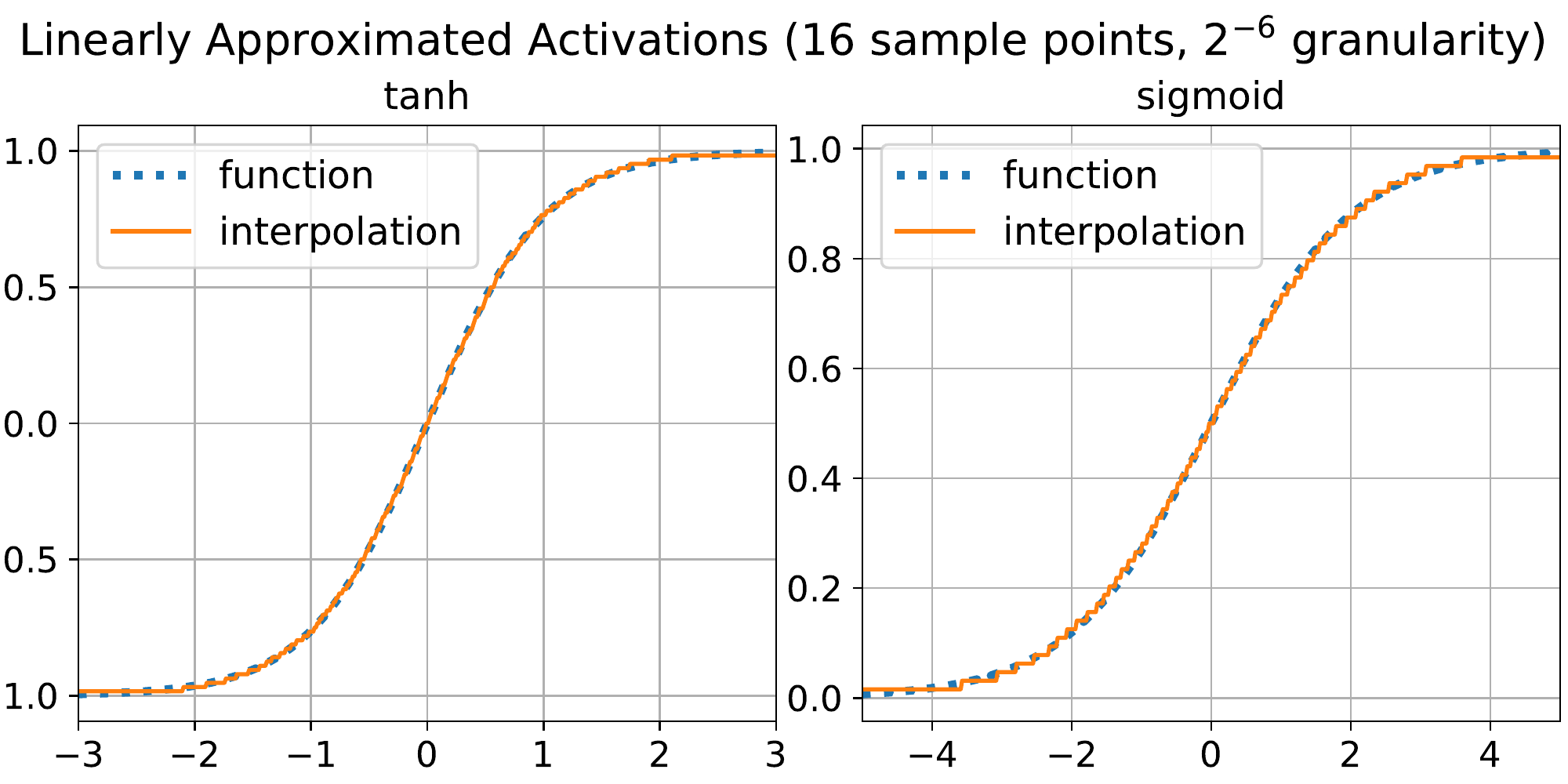}
	\caption{\label{f:Activations_Interpolated}
		Example for a linear interpolation of activations functions. Python-based case study, every value was rounded to a granularity of only $2^{-6}$ during all steps of the interpolation, only 16 sample points were used.
	}
\end{figure}

Other activations can be implemented in the future. One way of doing this is by either value-based look-up tables or by value-derivative-based look-up tables. For example, with 16 bit activation unit input values, one could use the 8 most significant bits to look up an interpolation base value and a first derivative, and then add the first derivative multiplied by the eight least significant input bits to the base value. With the Xilinx US+ architecture, both look-ups could be done at an approximate LUT cost of $2 \cdot 2^{8-6} \cdot 8 = 64$, the first two comes from two look-ups, the factor $2^{8-6}$ from the cost for looking up one bit for an $2^8$-deep address space and the eight comes from an assumed eight bits which are looked up for the base value and first derivative. The multiplication of two 8 bit values and final addition would cost approximately 70 LUTs, leaving a total of $\sim 140$ LUTs per activation unit to implement a relatively precise look-up with 256 sample points and linear interpolation in between.\footnote{The exact LUT cost depends on some details, but less than 200 is seen as reasonable assumption for many cases. Characteristics like base value and derivative saturation could be further exploited for an even decreased LUT requirement.} Figure \ref{f:Activations_Interpolated} shows a case study, where we linearly interpolated the comparably complicated tanh and sigmoid functions. To demonstrate the loose precision requirements, we used only sixteen sample points within the shown intervals, and rounded to a granularity of only $2^{-6}$ during all steps of the interpolation (i.e. the value and derivate samples themselves were rounded, the interpolation position was rounded and multiplication and addition results were rounded). This case study nicely demonstrates that even with only very low precision and very few sample points, it is possible to have a surprisingly accurate approximation of non-linear activations. In these cases, it would be possible to implement the look-up of both base value and derivative with approximately ten LUTs and the multiplication and addition with coarsely 50 LUTs. Given that there would usually be only a single activation unit at the end of a many-DSP pipeline, one can expect an extra utilization of much less than 10 LUTs per DSP even for a precise interpolation.

\section{Network creation toolkit}

\subsection{Layer synchronization considerations}

When connecting fully-connected layers to fully-connected layers, it is only reasonable to select a pipeline parallelization factor $P$ in the successor layer that equals the number of neuron units $N_\mathrm{NU}$ in the predecessor layer. By that, every input is used exactly when it is made available, and no extra buffering structures or delays are necessary.

When connecting layers with 'arbitrary' input and output schemes, there is no sense in starting the second layer earlier or later than necessary to run without interruption. This means that it is necessary to compute the minimum start delay that allows uninterrupted computation of the successor layer. Minimum start delay determination can easily be done if the output and input pattern of the involved layers are known, i.e. at which cycle which results/inputs are produced/needed. By element-wise subtraction of the 'needed' pattern from the 'available' pattern, one can take the largest positive value as minimum necessary delay for continuous computation.

However, with a given start delay, it might happen for some layer sequences that an input is already overwritten before it is needed for the last time.\footnote{For example the 2D convolution might have multiple load operations from its buffer memory and can therefore create such a situation.} These cases can be solved by the introduction of individual extra delays for all affected input values. The extra delay for each input value can be computed by adding the start delay to the input 'needed' scheme, adding $C - 1$ to the 'available' scheme, subtracting the latter from the former and then introducing the respective extra delay for all values which yield a positive result. This does \emph{not} increase the layer-to-layer latency.

Further delays that need to be taken into account are possible extra delays introduced by the application of activation functions and delay offsets between the layer enable signal and when that layer actually expects the first input. The latter do not influence the network latency itself, but need to be considered for asserting the layer enable signals at the right moments, which is automatically done in our toolkit.

In the special case of flattening, it is currently necessary to 'regularize' the fully-connected layer input, because the fully-connected layer has no buffer memory, and therefore it is necessary to ensure that the inputs are updated in the exact order in which the computations propagate through the pipeline. This requires an extra auxiliary layer resulting in one extra cycle delay. The same effect could in the future be obtained by re-ordering the inputs/weights of the fully-connected layer, such that the inputs which are available first after flattening are also required first. This was not yet implemented because the regularization approach is more general, and it was understood only lately that a mere re-ordering would be sufficient, if the fully-connected layer parallelization $P$ is set to the number of simultaneously produced result values from the last 2D layer.

\subsection{Network creation}

The Python-based toolkit for automated network creation and the VHDL library files can be obtained via email from the authors and is distributed as open source software. 
The starting point for the network implementation on FPGA is a trained Keras network. Supported network architectures consist of the previously described layers and activation functions. An arbitrary sequence of 2D multi-channeled layers \emph{can} be followed by flattening and an arbitrary sequence of fully-connected layers, or a fully-connected-only network can also be used.

Before the network can be implemented for the FGPA, it is necessary to specify/customize various design parameters. These include the precision (i.e. integer and fractional bits) of the input value and weight representation, which can even be customized on a layer-wise basis. Apart from specifying layer input and output bit widths, it is even possible to configure how intermediate values are treated, which occur for partial results being passed between DSPs and from the last DSP within a pipeline to a potential logic-based adder. Other parameters include pipelining and routing behavior, for example whether the relu activation is implemented in LUTs or FFs, if data flowing between DSPs is routed through the dedicated neighbor connections or the general fabric routing, or where registers are placed in arithmetic and logic pipelines.

Based on this information and the trained Keras network, the toolkit can be used for creating the VHDL network top file, auxiliary package file for configuration constants, network simulation testbench file and network data files for initializing control and weight memories from a Keras network. The VHDL files can finally be included in a project together with the network library files and are ready for use. Then, it is only necessary to load the respective weights and controller data into the memories during execution, after which inference is possible.

\section{Results}

Given our target design parameters, i.e. \SI{40}{MHz} data frequency, at most \SI{600}{MHz} to \SI{900}{MHz} processing frequency and the Xilinx XCVU9P FPGA, we considered a possible network size of at most multiple $10^4$ MACs, to constrain ourselves to reasonable layer and network sizes for further studies. 
Typical deep neural networks in use by ATLAS start at a few $10^3$ MACs for fully-connected networks aiming at tagging  $W$-Bosons and Top-Quarks~\cite{ATL-PHYS-PUB-2017-004}, but can easily reach the order of several  $10^7$ MACs for e.g. quark/gluon identification using convolutional networks on calorimeter images~\cite{ATL-PHYS-PUB-2017-017}, as they have not been optimized for low resource usage.

For any design, timing closure was characterized by the 'worst negative slack' (WNS), which can be thought of as difference between the signal propagation time estimated by Vivado and the target clock period. Timing characteristics apart from the WNS were met for all implemented designs. All designs were implemented with the 'out of context synthesis'. The bit lengths of the layer input values and results were set to 14 (6.8), and where occurring weights were set to 10 (2.8) bits length.\footnote{If any, we observed less than 1\% accuracy degradation in all of our MNIST digit recognition test networks with that precision choice. Fewer or more bits might be necessary for satisfying accuracy with other networks, with an approximately linear effect on the utilization when both bit widths are scaled simultaneously.}

\subsection{Isolated layers}

All implementations were done with a target clock period of \SI{1.563}{ns}, reflecting the $f_\mathrm{P} = \SI{640}{MHz}$ processing frequency that are ideally achievable even with speed grade 1 US+ DSPs, which corresponds to $C = 16$ for an assumed data rate of \SI{40}{MHz}, as in the ATLAS experiment. Therefrom, we derived the minimum processing clock period $T_\mathrm{P,min} = T_\mathrm{P} - T_\mathrm{WNS}$ of a design. Designs which failed with a large WNS might show better results for larger $T_\mathrm{P}$, because for better-suited $T_\mathrm{P}$, Vivado can typically reach better $T_\mathrm{P,min}$.

\begin{figure}
	\includegraphics[width = \textwidth]{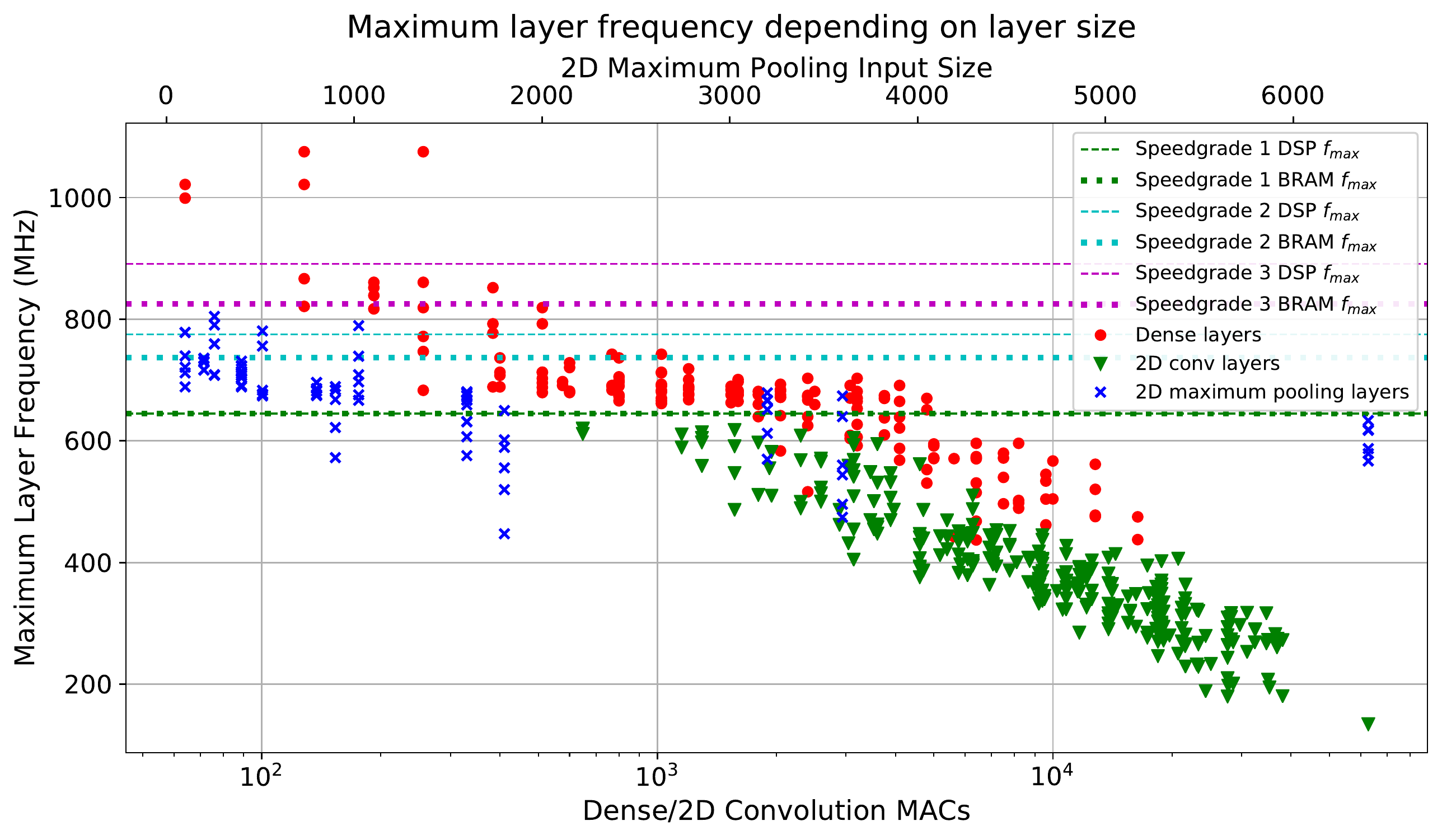}
	\caption{\label{f:Layer_SizevFreq}
		Layer size versus maximum frequency. With maximum DSP and BRAM frequencies depending on speed grade. The clustering between  \SI{640}{MHz} and \SI{750}{MHz} probably is an artifact of the target frequency of \SI{640}{MHz}. (Designs which reach timing closure are barely optimized further, and designs which initially just fail can often be optimized to meet the timing. Only too large designs fall completely below the target frequency.)
	}
\end{figure}

\paragraph{Fully-connected layers}

We implemented fully-connected layers for $N_\mathrm{I}, N_\mathrm{N} \in \lbrace 8, 16, 24,\allowbreak 32,\allowbreak 50,\allowbreak 64,\allowbreak 75,\allowbreak 100,\allowbreak 128 \rbrace$, \mbox{$C \in \lbrace 10, 16 \rbrace$}. As expected, the DSP utilization represents what was designed for, i.e. $N_\mathrm{I} \cdot N_\mathrm{NU}$. BRAM utilization is at most $0.5 \cdot \ceil[\big]{ \frac{N_\mathrm{I}}{P} } \cdot N_\mathrm{NU} \cdot \ceil[\big]{ \frac{P}{3} } + 0.5$, since for $P > 3$ and 10 bit weights, multiple BRAM (sub-)units are necessary for most stage weight memories. For larger layers, this corresponds to approximately one BRAM per 5 DSPs, which is tolerable given the resource ratios in the US+ device family. For all of these designs, at most 4 LUTs and 23 FFs were required per DSP, with less than 10 FFs per DSP in almost 75\% of the designs. Larger designs tended to have lower per-DSP LUT and FF utilization, due to positive scaling effects. Compared to $\sim 170$ LUTs and $\sim 340$ FFs per DSP in the XCVU9P FPGA, both values are negligible. Layer size in terms of MACs against maximum frequency is shown in figure \ref{f:Layer_SizevFreq}. Depending on the device choice, it is possible to have up to multiple thousand MACs in a layer, before the implemented design rather than the device specification becomes limiting.


\paragraph{2D convolutional layers}
\label{ssp:Results_Conv}

The exact structure and therefore resource utilization of 2D convolutional layers depends strongly on the architectural parameters. To give a good overview, we implemented for some hand-picked parameters, to ensure all architectural features are covered at least once, and did a grid implementation for $H_\mathrm{I} \times W_\mathrm{I} \in \lbrace (10, 10), (15, 15), (16, 16), (20, 20), (25, 25) \rbrace$, $I_\mathrm{D} \in \lbrace 1, 2, 4, 6 \rbrace$, $H_\mathrm{K} \times W_\mathrm{K} \times N_\mathrm{K} \in \lbrace (2, 2), (3, 3), (4, 4) \rbrace \times \lbrace {2, 4, 6} \rbrace$, $C \in \lbrace 10, 16 \rbrace$ with a veto for these if more than 3000 DSPs (i.e. almost half of the XCVU9P DSPs) are needed for the given single layer. By design, no BRAM is used. The number of DSPs needed is $N_\mathrm{RU} \cdot W_\mathrm{O} \cdot H_\mathrm{K} \cdot W_\mathrm{K} \cdot D_\mathrm{I}$. For the ratios of LUTs and FFs to DSPs, we achieved 6 to 61 (with 75\% of the designs below 25, and 95\% below 40) and 10 to 77 (with 75\% of the designs below 40 and 90\% below 50), respectively, which is still a good result given the available resource ratios. Again, larger layers tend to require fewer LUTs and FFs per DSP. A plot of layer size in terms of MACs against maximum frequency is shown in figure \ref{f:Layer_SizevFreq}. It turned out that for comparable size, convolutional layers typically reach lower maximum frequencies than fully-connected layers. This is a result of the optimization for resource utilization rather than timing at a time when the resource consumption was not yet known. Critical signal paths could already be identified, and we expect to be able to make it possible to trade a higher frequency for more utilization in the future.


\paragraph{2D max-pooling layers}

Test implementations of layers included $H_\mathrm{P} \times W_\mathrm{P} \in \lbrace (2, 2), (3, 3), (4, 4) \rbrace$,
$H_\mathrm{I} \times W_\mathrm{I} \in \lbrace (10, 10), (16, 16), (20, 20), (30, 30), (40, 40) \rbrace$, $D_\mathrm{I} \in \lbrace 1, 2, 4 \rbrace$, $C \in \lbrace 10, 16 \rbrace$, padding 'same'. Per input value, between 5 and 16 LUTs were required, with 70\% of the values below 8, and 18 to 29 FFs, with 65\% of the values below 20. Hence, pooling layer utilization is negligible for computationally reasonable input sizes, with even only up to 52k LUTs and 130k FFs needed for as many as 6400 inputs.\footnote{That many inputs would already be difficult to handle with convolutional layers, depending on the exact layer parameters.} Layer size in terms of inputs against maximum frequency is shown in figure~\ref{f:Layer_SizevFreq}. Small layers met the target frequency of \SI{640}{MHz} without any problems, some larger designs fell below that line, but not significantly. Future alternative designs could yield increased frequencies, but this was not yet studied, as an improved implementation of 2D convolutions appears more urgent.


\subsection{Example networks}
\label{ss:Results_Networks}

We tested four basic network architectures, where in the following we use I, C, P, F and D for 2D input, 2D convolutional, 2D pooling, flattening and fully-connected layers respectively. Architecture Arc\textsubscript{A} has the layer sequence \mbox{I-C-P-F-D-D}, Arc\textsubscript{B} has \mbox{I-C-P-C-F-D-D}, Arc\textsubscript{C} has \mbox{I-C-P-C-F-D-D-D} and Arc\textsubscript{D} has \mbox{I-C-C-C-F-D-D-D}. For each architecture, we trained various networks with varying layer parameters for the MNIST digit recognition task. The last fully-connected layer always had 10 neurons to classify the ten digits. A range of values for $C$ was also scanned for each network, where as data frequency $f_\mathrm{D}$ we chose \SI{40}{MHz}, and as (target) processing frequency $C \cdot f_\mathrm{D}$. We used a precision choice of 6.8 and 2.8 for values and weights, for which at most a sub-percent order of classification accuracy loss was observed, compared to CPU inference using float32 as data type. Some of the implementation results are shown in table \ref{t:ResNets}.

\begin{table}
	\caption{\label{t:ResNets}
		Implementation results for example networks trained for the MNIST digit recognition task. Activation relu used for all but the last layer, which had a 'linear' activation and always had $N_\mathrm{N} = 10$ neurons. The single-channel input has a size of $14 \times 14$, if not stated otherwise. WNS was left out where no WNS occurred, i.e. timing was met for these designs. For convolutional layers, the kernel shape $(H_\mathrm{K} \times W_\mathrm{K} \times N_\mathrm{K})$ is specified, for pooling layers the pooling area $(H_\mathrm{P} \times W_\mathrm{P})$ and for fully-connected layers the number of neurons $N_\mathrm{N}$ is specified. The DSP efficiency refers to the relative amount of non-idling DSP cycles. For reference, the Xilinx UltraScale+ XCVU9P target FPGA features 6840 DSPs, 2160 BRAMs, approximately 1.2 million LUTs and twice as many FFs.
	}
	
	\centering	

        \begin{ADLactivate}
	\begin{tabular}{l|c|ccc|cc}
		Architecture (see text) & MACs & $T_\mathrm{P}$ & WNS & latency & $N_\mathrm{LUT}$ & $N_\mathrm{FF}$
		\\
		{\scriptsize(layer information)} & (DSP eff.) & (ns) & (ns) & (cycles) & $N_\mathrm{DSP}$ & $N_\mathrm{BRAM}$
		\\
		
		\hline
		
		Arc\textsubscript{A1} ($C = 16$) (input $(7 \times 7)$) & 334 & 1.562 & - & 56 & 1793 & 3571 \\
		{\scriptsize $(2\times 2 \times1)$-$(2\times 2)$-$10$} & (0.485) & & & & 43 & 10.5\\

                \arrayrulecolor{gray}
                \cdashline{1-7}
          
		Arc\textsubscript{A2} ($C = 14$) & 1089 & 1.786 & - & 60 & 5060 & 9706 \\
		{\scriptsize $(2 \times 2 \times 1)$-$(2\times 2)$-$7$} & (0.630) & & & & 108 & 17 \\

                \cdashline{1-7}
          
		Arc\textsubscript{A3} ($C = 14$) (input $(7 \times 7)$) & 1024 & 1.786 & - & 57 & 3051 & 5654 \\
		{\scriptsize $(2 \times 2 \times 3)$-$(2 \times 2)$-$16)$} & (0.620) & & & & 118 & 19 \\

                \cdashline{1-7}
          
		Arc\textsubscript{A4} ($C = 13$) & 3188 & 1.923 & - & 63 & 8689 & 16219 \\
		{\scriptsize $(2\times2\times2)$-$(2\times2)$-$17$)} & (0.774) & & & & 317 & 54.5 \\
		
                \cdashline{1-7}

		Arc\textsubscript{A5} ($C = 13$) & 7854 & 1.923 & - & 68 & 15567 & 28450 \\
		{\scriptsize $(2\times2\times4)$-$(2\times2)$-$25$} & (0.967) & & & & 625 & 93.5 \\
		
                \cdashline{1-7}

		Arc\textsubscript{A6} ($C = 11$) & 12884 & 2.273 & - & 68 & 20962 & 34711 \\
		{\scriptsize $(3\times3\times4)$-$(2\times2)$-$50$} & (0.894) & & & & 1310 & 166 \\
		
                \cdashline{1-7}

		Arc\textsubscript{B1} ($C = 12$) & 8858 & 2.083 & - & 76 & 18587 & 32886 \\
		{\scriptsize $(2\times2\times4)$-$(2\times2)$-$(2\times2\times4)$-$25$} & (0.812) & & & & 909 & 99.5 \\
		
                \cdashline{1-7}

		Arc\textsubscript{B1} ($C = 16$) & 8858 & 2.083 & - & 87 & 17205 & 32760 \\
		{\scriptsize $(2\times2\times4)$-$(2\times2)$-$(2\times2\times4)$-$25$} & (0.812) & & & & 713 & 71.5 \\
		
                \cdashline{1-7}

		Arc\textsubscript{B3} ($C = 11$) & 11362 & 2.273 & - & 79 & 28383 & 47140 \\
		{\scriptsize $(2\times2\times6)$-$(2\times2)$-$(2\times2\times4)$-$25$}& (0.792) & & & & 1305 & 102.5 \\

                \arrayrulecolor{black}
		\hline
		
		Arc\textsubscript{B2} ($C = 10$) & 15610 & 2.500 & -0.134 & 84 & 40998 & 69333 \\
		{\scriptsize $(3\times3\times6)$-$(2\times2)$-$(3\times3\times6)$-25} & (0.855) & & & & 1825 & 68 \\

                \arrayrulecolor{gray}
                \cdashline{1-7}

		Arc\textsubscript{B3} ($C = 16$) & 11362 & 1.562 & -0.014 & 93 & 26006 & 45065 \\
		{\scriptsize $(2\times2\times6)$-$(2\times2)$-$(2\times2\times4)$-25} & (0.825) & & & & 861 & 71.5 \\
		
                \cdashline{1-7}

		Arc\textsubscript{C1} ($C = 8$) & 24076 & 3.125 & -0.045 & 93 & 37528 & 61388 \\
		{\scriptsize $(3\times3\times6)$-$(2\times2)$-$(2\times2\times8)$-50-25} & (0.934) & & & & 3222 & 338.5 \\
		
                \cdashline{1-7}

		Arc\textsubscript{D5} ($C = 9$) & 26120 & 2.778 & -0.060 & 86 & 32592 & 51865 \\
		{\scriptsize $(2\times2\times4)-(2\times2\times2)-(2\times2\times2)-50-25$} & (0.928) & & & & 3128 & 353 \\

	\end{tabular}
        \end{ADLactivate}
\end{table}

\begin{figure}
	\centering
	\includegraphics[width=.8\textwidth]{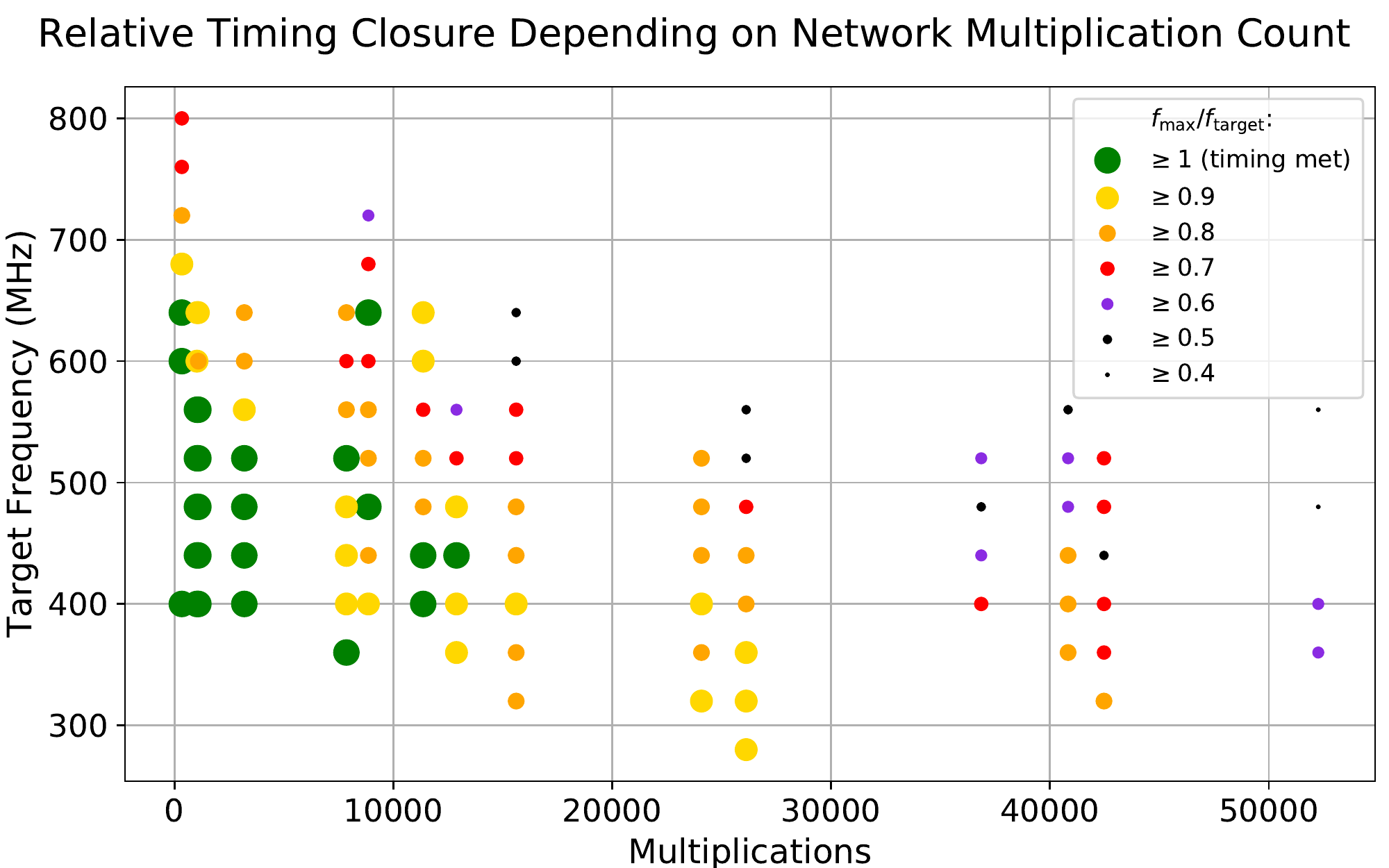}
	\caption{\label{f:Networks_FreqClosure}
          Network frequency closure depending on network size and target processing frequency. 'Frequency closure' is defined as  the ratio between the targeted processing frequency and the maximum processing frequency according to Vivado. Note that especially for the larger networks, there is a lower limit for $C$ due to the amount of available resources, and large values for $C$ are excluded due to the difficulty to reach such high frequencies with those designs.
	}
\end{figure}

\begin{figure}
	\centering
	\begin{subfigure}[t]{.48\textwidth}
		\includegraphics[width = \textwidth]{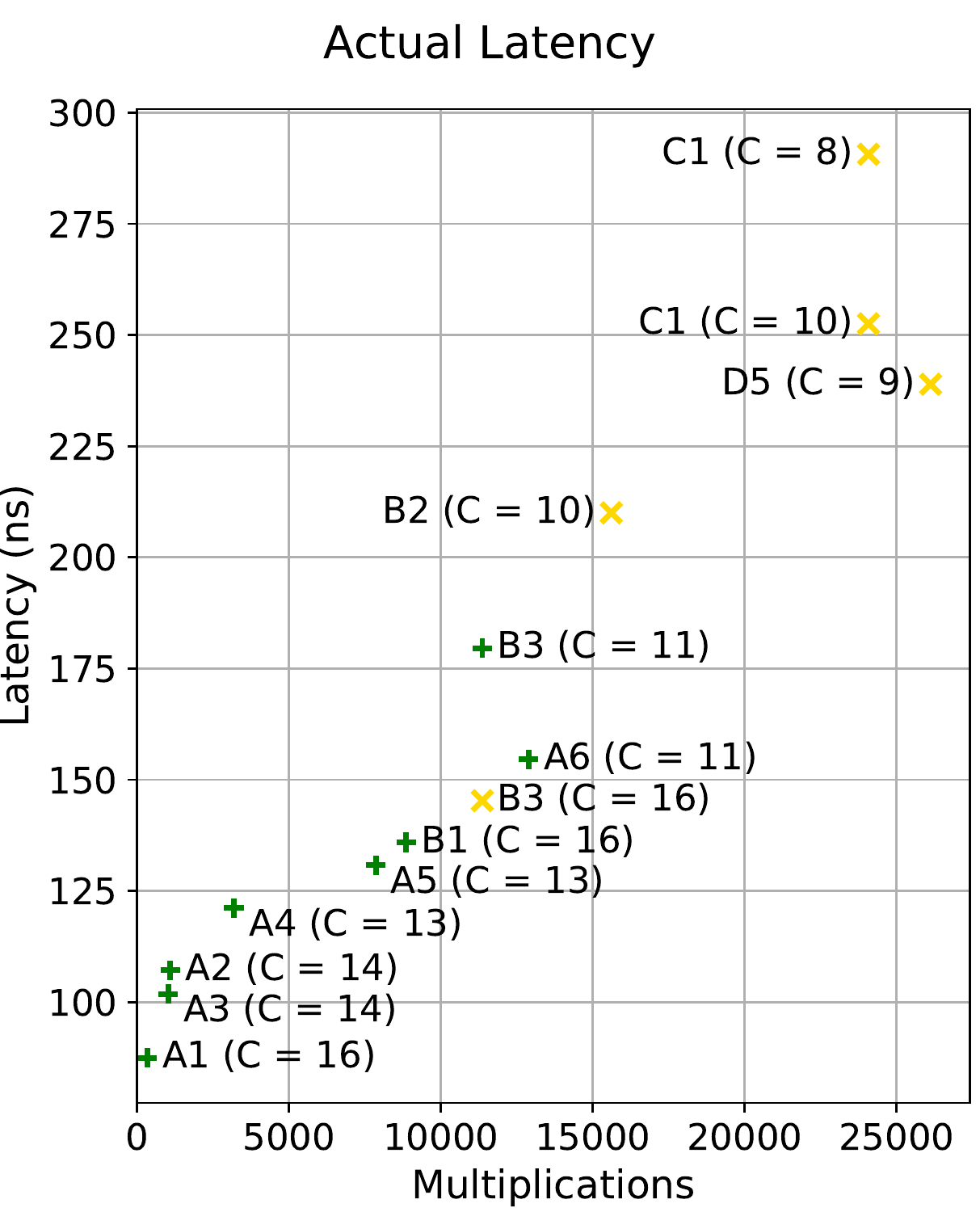}	
		\caption{\label{f:Networks_Latencies}
			Network latency versus size (in terms of MACs) for networks which achieved timing closure (plus indicator) or failed only slightly (cross indicator).
		}
	\end{subfigure}
	\hfill
	\begin{subfigure}[t]{.48\textwidth}
		\includegraphics[width = \textwidth]{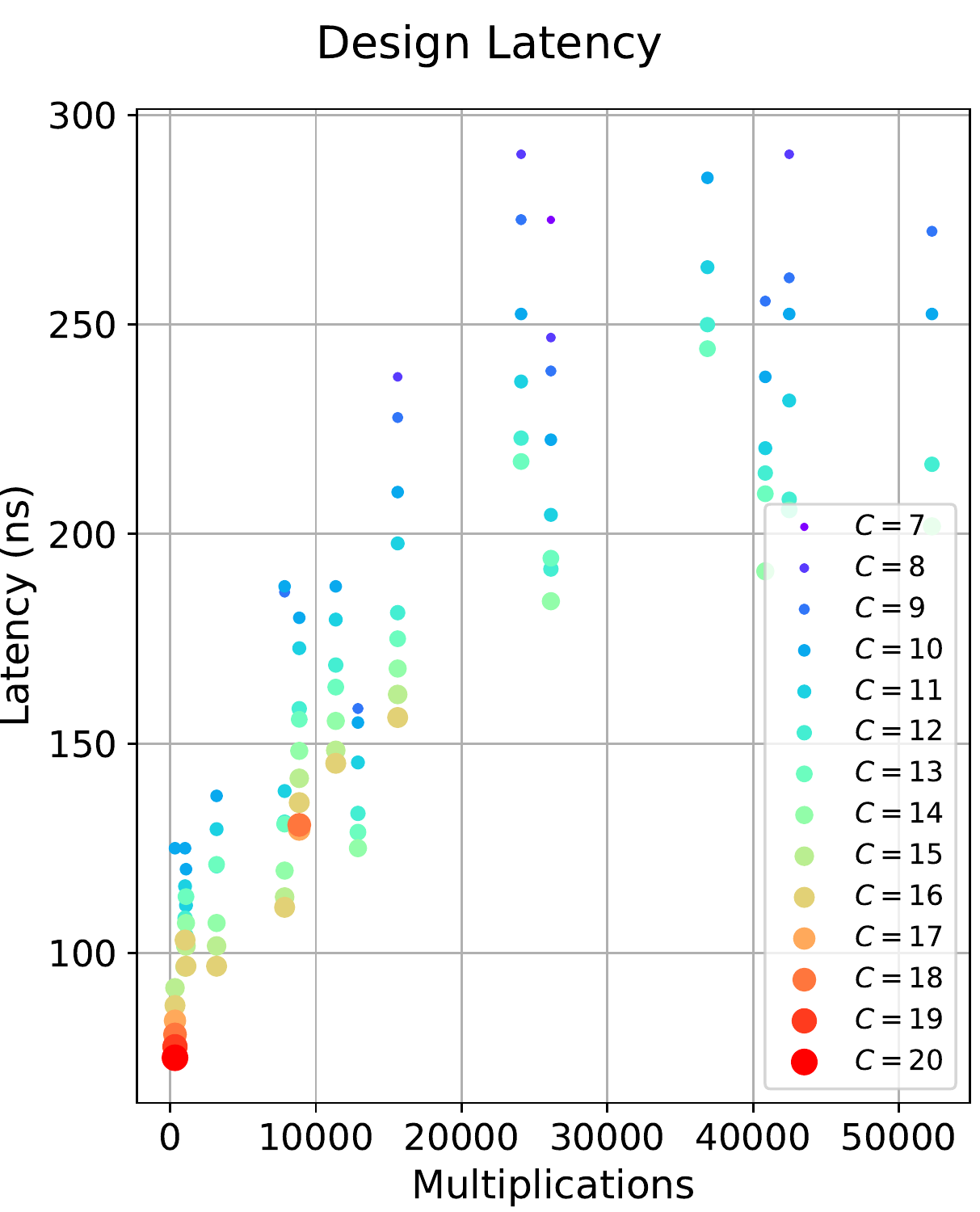}	
		\caption{\label{f:Networks_Latencies_C}
			Network latency depending on $C$ if it is assumed that timing closure can be achieved, with networks and values for $C$ corresponding to the data from figure \ref{f:Networks_FreqClosure}.
		}
	\end{subfigure}
	
	\caption{Timing characteristics for some of the MNIST example networks.}
	
\end{figure}

For the given precision choice, between 7 and 50 LUTs were needed per DSP (among all networks, including those which completely failed timing and are not shown in table \ref{t:ResNets}), with the tendency of less LUTs per DSP for larger networks, and less than 30 LUTs per DSP in 83\% of the cases. Similarly, between 12 and 90 FFs were needed per DSP, with less than 50 FFs in 80\% of the cases, again with a tendency for smaller values in larger networks. The absolute BRAM utilization depends largely on the number of inputs and neurons in the fully-connected layers, but does never exceed 20\% of the absolute DSP utilization, and is significantly lower in most cases.

The larger networks ($>15\mathrm{k}$ MACs) did not yet meet the timing requirements, which could already be expected based on the results for the individual layers. For all of the networks below that size, at least one value for $C$ was found for which the network met the timing. The findings regarding the relation between network size and timing closure are also shown in figure \ref{f:Networks_FreqClosure}. One interesting observation is that the relation between timing closure and $C$ is not nearly 'continuous', i.e. there might be designs with a large timing violation for one choice of $C$, and little violation or even timing closure for neighboring values of $C$, or vice-versa. This is no surprise, as $C$ has a large influence on the structure of the design, but it shows that it might be necessary to explore different settings to find a working design for medium-sized networks. However, it can still be seen that in tendency, timing closure is more difficult to reach for larger networks. For the future, it is expected that frequency improvements for the individual layers (especially for the convolutional layer, see section \ref{ssp:Results_Conv}) will make it possible to reach timing closure at least for networks which failed only closely. If then layer-to-layer paths become critical, it is an option to add extra register stages between layers to relieve the placement and routing efforts.

Network latencies (both in terms of real time latency and cycles) are put into relation to network size in figure \ref{f:Networks_Latencies}. The two dominant factors to the network latency are the number of layers and the processing frequency. For the four-layer networks with architecture Arc\textsubscript{A}, a latency of $\sim \SI{100}{ns}$ is obtained, after which \emph{all} results are available. The larger networks which just failed timing closure required up to $\sim \SI{300}{ns}$ to produce all results. Figure \ref{f:Networks_Latencies_C} shows what latencies would be obtained depending on $C$ if timing closure was always achieved for networks with up to $\sim \SI{50}{k}$ MACs, which especially shows that with increasing $C$, a significant latency reduction would become possible. It is expected that with future design improvements, the network latencies will approximately remain the same (or even improve) in terms of cycles, but improve in terms of real time latency due to frequency improvements, which would be represented by figure \ref{f:Networks_Latencies_C}.

Regarding the DSP efficiency, our design is not always able to adapt efficiently for very small networks, as expected. For larger networks, where the intended design works well, efficiencies of more than 90\% can be reached. Please note that this is not yet completely visible in table \ref{t:ResNets}, because the networks which are shown there are still relatively small compared to the theoretical maximum network size, although an on average increased DSP efficiency can also be observed there.

We also switched the relu activation implementation between LUT- or FF-based for some designs, but did not see a consistent effect on timing, only resource usage changed accordingly between LUTs and FFs. Vivado might not always be able to make use of the improved placement capabilities that come from the extra register layer. Additionally, we implemented one layer for significantly increased precision, and observed an almost linearly increased utilization and slightly worse timing characteristics, as expected.

\section{Summary and outlook}

We were able to develop a neural network implementation framework that is suitable for use within detector triggers, provides a substantial inference performance even at incoming data rates of many MHz, and requires only tens to few hundreds of nanoseconds for inference for reasonable network sizes by specifically designing neural network layers for the special constraints of the trigger environment. By embedding our developments into an easy-to-use toolkit, the overhead of implementing efficient, high-performance neural networks within trigger systems with such constraints has been greatly reduced, as no expert knowledge about FPGA implementation and neural networks is needed any more. Now, implementing a neural network on trigger FPGAs, for example in the ATLAS detector, can be as simple as running our script on an already trained Keras network.

In the future, we intend to further improve and extend our work, with already a lot of useful features in our current scope. Among these are various changes to the current layer implementations, which target network latency and maximum frequency, but also functional extensions such as support for neuron bias, new activation functions and new layer types, such as transposed 2D convolutions for upsampling. We also have toolkit extensions planned, such as layer emulation for faster network benchmarking and automated precision recommendations. Further in the future, we see potential to significantly increase the overall performance, for example by implementing support for sparse weights, but also  by further improving the efficiency of the FPGA resource utilization.

\bibliography{mybib}{}
\bibliographystyle{ieeetr}

\newpage
\appendix

\section{Appendix}

\subsection{Details on irregular 2D convolution implementations}
\label{ss:Conv_Irreg}

\begin{figure}[!b]
  \centering
  \includegraphics[width=0.5\textwidth]{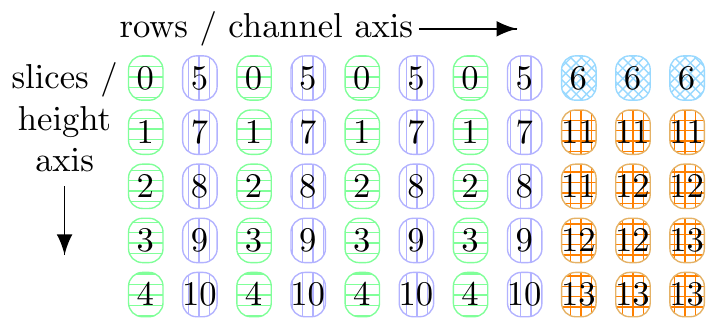}
  \caption{\label{fs:Conv_Irreg_Allocs}
    'Worst case' example of an irregular allocation scheme, occurring for $C = 15$, $D_\mathrm{O} = 11$, $H_\mathrm{O} = 19$, cropped to the irregular slices. The number within the rectangle refers to which RU processes that row. Green (horizontal lines) for ssl-RUs, purple (vertical lines) for bslc-RUs, light blue (crosshatch) for bsli-RUs, orange (grid) for msl-RUs.
  }
\end{figure}

In the irregular case of the convolutional layer, the number of slices $N_\mathrm{Sl}$ exceeds the number of complete slices processable by the row units $N_\mathrm{Sl,compl} = N_\mathrm{RU,Sl} \cdot N_\mathrm{RU}$, which is when the remainder RU cycles need to be used for maintaining a high computational efficiency. Also see figure \ref{fs:Conv_Irreg_Allocs} for the following. In this case, RUs are not distinguished by how many complete slices they process (i.e. 'long' and 'short' RUs), but instead we introduce five categories of RUs: 'single-slice' (ssl), 'complete bi-slice' (bslc), 'incomplete bi-slice' (bsli), 'multi-slice' (msl), and 'free' row units. All RUs stick to the regular pattern from before until the remainder cycles are reached: For example for the layer from figure \ref{fs:Conv_Irreg_Allocs} ($D_\mathrm{O} = 11$, $H_\mathrm{O} = 19$ and $C = 15$, resulting in $N_\mathrm{RU} = 14$), RU $i$ would start computing the outputs of result slice $i$, until all results of that slice were computed after eleven cycles, and the irregular scheme displayed there only describes the last four cycles and the then remaining five result slices.

	
	
	

The ssl-RUs then stick to the original slice step pattern, i.e. they step from slice $i$ to $i + N_\mathrm{RU}$ (with $i$ being the respective slice they were just computing) and compute output rows of that slice. Hence, there are as many ssl-RUs as remaining slices. However, in this case it is not possible for the ssl-RUs to compute the entire output slice anymore, which is why other RUs also need to contribute to these output slices. The bsl-RUs only process rows of one extra slice during their remainder cycles, but that extra slice does \emph{not} stick to the usual slice step width of $N_\mathrm{RU}$, as the respective outputs do not exist. The difference between bslc-RUs and bsli-RUs is that the former use all of their remainder cycles, while the latter do not. bslc-RUs are added to the remainder slices until no further of them can be added, which is when only less than $C \bmod D_\mathrm{O}$ rows are left per remainder slice. For sticking to the initial 'top channel to bottom channel' result pattern, ssl- and bslc-RUs compute output rows in an interleaved pattern. Then, it is determined how many extra bsli-RUs can be added at most, such that the remaining RUs can still cover all then remaining rows. If there are still rows uncovered after adding the bsli-RUs, msl-RUs are introduced, which process rows of more than one of the irregular output slices. There may be RUs which do not need to contribute to the irregular result slices at all, these are considered 'free'. There cannot be free and msl RUs at the same time. bsli- and msl-RUs were not yet interleaved with the rest due to complexity reasons. In the future, improvements might be made to this scheme, which could aim at using all of the RUs more evenly for the irregular output slices, in order to produce the results even faster and thereby potentially reduce the layer-to-layer data propagation time.

\begin{figure}
  \centering
  \includegraphics[width=0.75\textwidth]{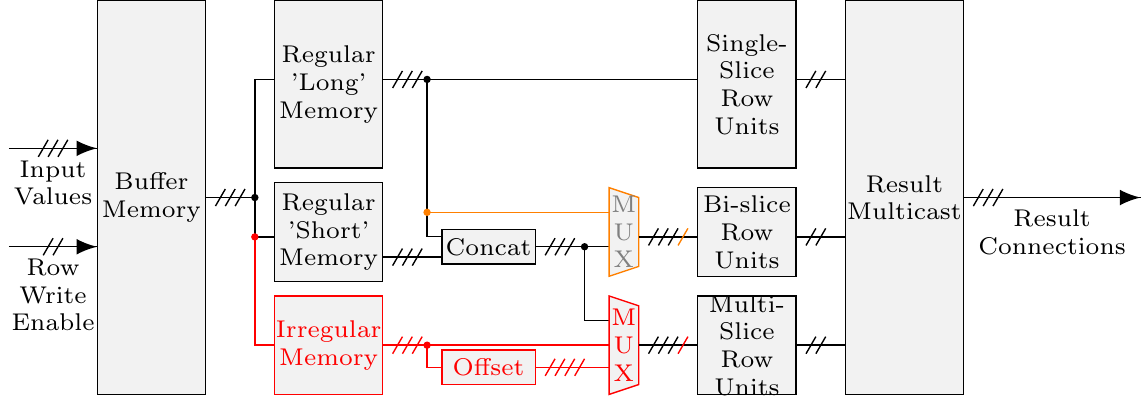}
  \caption{\label{fs:Conv_Irreg_Structure}
    Convolutional layer schematic. The regular/'long' memory corresponds to the 'long' memory from the regular case, ssl-RUs correspond to 'long' RUs from the regular case. bsl-RUs correspond to 'short' RUs during regular cycles, and have their inputs switched to the ssl-RU inputs during irregular cycles. Msl-RUs correspond to 'short' RUs during regular cycles, and have their inputs switched to an extra memory, with a cycle- and unit-dependent extra offset of 1 input slice.
  }
\end{figure}

Regarding the code structure, some extensions are necessary, see figure \ref{fs:Conv_Irreg_Structure}. ssl-RUs and 'free' RUs correspond to the 'long' and 'short' RUs from the regular case regarding how they are wired. The bsl-RUs and msl-RUs correspond to 'short' RUs during the regular cycles. During the irregular cycles, the bsl-RUs have their inputs multiplexed back to the ssl-RU inputs, as they also require these then. The msl-RUs need their own working memory for the irregular phase, as they require inputs from multiple slices over time, which cannot be reflected by using the ssl-RU working memory. Additionally, msl-RUs typically do not step to the next slice synchronously (see figure \ref{fs:Conv_Irreg_Allocs}, RU 11 steps after three cycles, RU 12 after two, RU 13 after one), which means that it is possibly necessary to introduce an extra slice offset of one during certain cycles, until the entire memory follows up. Hence, multi-slice units have their inputs not only multiplexed between regular and irregular, but also between irregular without and with an extra offset of one slice, which requires one multiplexer for each msl-RU. As in the regular case, ssl- and 'free' RUs each have one set of weight memories for all of these RUs. bsl-RUs share their weight memories only if they are also working on the same channels during the irregular cycles. Each msl-RU has its own set of weight memories.

\end{document}